\documentclass[10pt, conference, compsocconf]{IEEEtran}

\newcommand{\oneImage}{3.2in}
\newcommand{\oneHalfImages}{2.12in}
\newcommand{\twoImages}{1.6in}

\usepackage{amsmath}

\usepackage{algorithm}
\usepackage{algpseudocode}
 
\algblock[Switch]{Switch}{end}
\algloop[Case]{Case}
\algblock[Choose]{choose}{end}
\algblock[Do]{do}{end}
\algblock[Phase]{Phase}{end}

\usepackage[
  bookmarks=false,
  pdfdisplaydoctitle=true
  pdftitle={A Scalable Distributed Architecture for Network- and QoS-aware Service Composition},
  pdfauthor={Adrian Klein},
 ]{hyperref}

\usepackage[pdftex]{graphicx}
\usepackage[justification=centering]{caption}

\begin{document}

\title{	A Scalable Distributed Architecture for\\Network- and QoS-aware Service Composition {\normalsize (IJAWS 2012$^{\ref{note1}})$} }

\author{
	\IEEEauthorblockN{Adrian Klein}
	\IEEEauthorblockA{University of Tokyo\\
		Tokyo, Japan\\
		adrian@nii.ac.jp}
\and
	\IEEEauthorblockN{Fuyuki Ishikawa}
	\IEEEauthorblockA{GRACE Center and\\
		National Institute of Informatics\\
		Tokyo, Japan\\
		f-ishikawa@nii.ac.jp}
\and
	\IEEEauthorblockN{Shinichi Honiden}
	\IEEEauthorblockA{University of Tokyo and\\
		National Institute of Informatics\\
		Tokyo, Japan\\
		honiden@nii.ac.jp}
}

\maketitle 

\setcounter{footnote}{0} 

\begin{abstract} 
Service-Oriented Computing (SOC) enables the composition of loosely coupled service agents
	provided with varying Quality of Service (QoS) levels,
		effectively forming a multiagent system (MAS).
Selecting a (near-)optimal set of services for a composition in terms of QoS is crucial
	when many functionally equivalent services are available.
As the number of distributed services, especially in the cloud, is rising rapidly,
	the impact of the network on the QoS keeps increasing.
Despite this and opposed to most MAS approaches, current service approaches depend on a centralized architecture which cannot adapt to the network.
Thus, we propose a scalable distributed architecture composed of a flexible number of distributed control nodes.
Our architecture requires no changes to existing services and 
	adapts from a centralized to a completely distributed realization	by adding control nodes as needed.
Also, we propose an extended QoS aggregation algorithm that allows to accurately estimate network QoS.
Finally, we evaluate the benefits and optimality of our architecture in a distributed environment.
{\small (IJAWS 2012\footnote{\label{note1}This paper has been accepted and published at the 3rd International Joint Agent Workshop and Symposium (IJAWS) 2012, Kakegawa, Shizuoka, Japan, October 2012.
This version has been copy-edited for publication at arXiv.org, but left unchanged besides.
Refer to \url{www.adrianobits.de} for the original submission in the IJAWS format.})}
\end{abstract}


\IEEEpeerreviewmaketitle

\section{Introduction}
\noindent Service-Oriented Computing (SOC) enables the composition of service agents
	in a loosely coupled way by realizing many ideas from the research of multiagent systems (MAS).
Services can be thought of as specialized agents only allowing access through their published interfaces.
SOC requires the modeling of autonomous and heterogeneous service components which form a MAS \cite{Huhns2005}.
The value of SOC is achieved by enabling rapid and easy composition of services with low costs \cite{Papazoglou2003}.

\subsection{QoS-aware Service Composition}
\noindent For service compositions, functional and non-functional requirements \cite{Sullivan2002} have to be considered.
The latter are specified by Quality of Service (QoS) attributes and are especially important
	when many functionally equivalent services are available.
A composition should be optimal in regards to the user's QoS preferences and constraints.
The QoS of a composition is the aggregated QoS of its services according to workflow patterns \cite{Jaeger2004},
	given each service's Service Level Agreement (SLA).

\noindent Thus, current approaches only consider the QoS of services themselves and ignore the QoS of the network.
One reason is that on a small scale services might be executed in a local network
	where network QoS is not significant.
With a growing distribution of services, this is no longer true.
Finally, the common opinion is that the provider of a service has to take the network into account in his SLA.
This is not trivial, as, in general, response times vary a lot depending on the user's location \cite{Zheng2010},
	making it hard for the provider to predict what kind of network QoS his users will experience.
Thus, the current practice becomes less accurate,
	as the number of distributed services keeps rising, deployed in locations around the world.
Therefore, we think it is essential to develop approaches tackling service composition in a network-aware manner
	to reduce the burden for providers to supply universal SLAs, and to improve the QoS for users.
  
\subsection{Network Delay}
\begin{figure}[htb!]
	\centering
	\includegraphics[width=\oneImage]{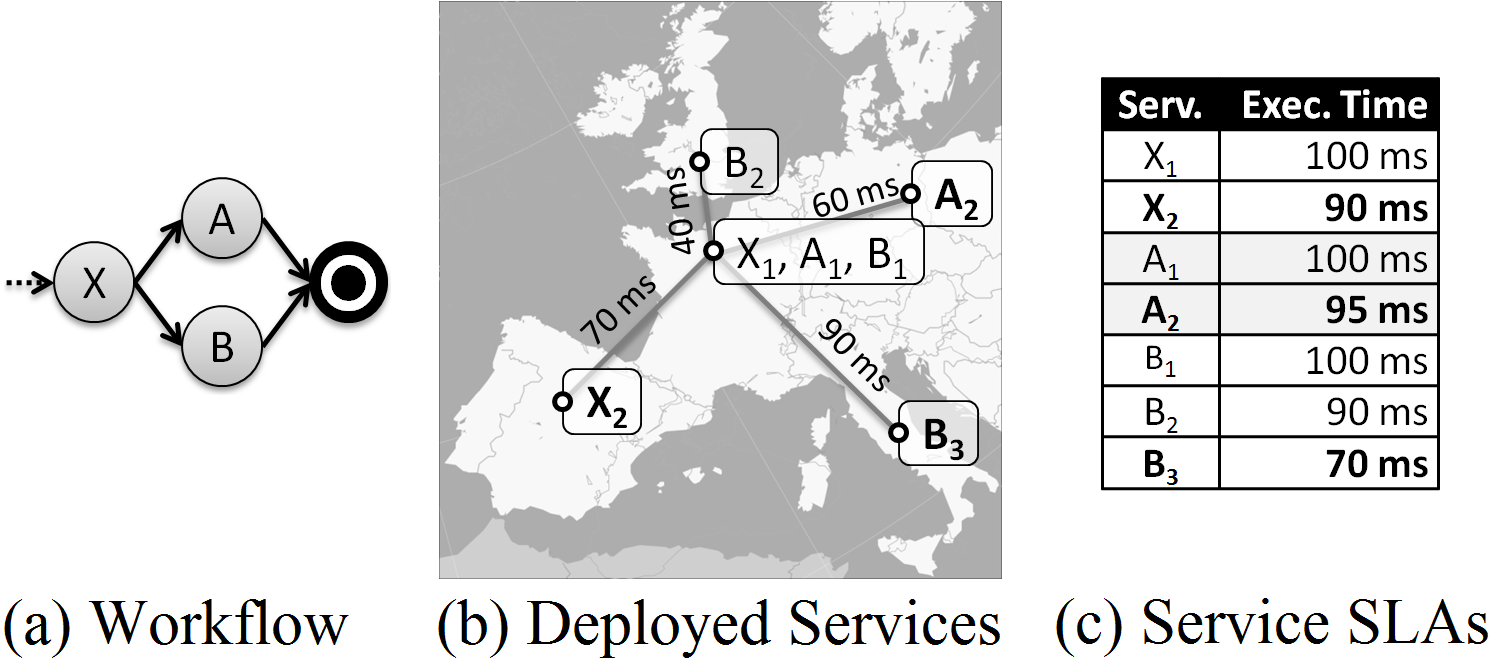}
	\caption{Distributed Deployment Example}
	\label{fig:example1}
\end{figure}

\noindent The example in Fig.~\ref{fig:example1} illustrates the necessity of a network-aware approach.
Consider the abstract workflow depicted in Fig.~\ref{fig:example1}(a),
	the corresponding concrete services ($X_1$, $X_2$, $A_1$, etc.) where $X_1$ performs task $X$, etc.
		and the execution times conforming to Fig.~\ref{fig:example1}(c).
We can see the deployment of the services and the network delays between the different deployment locations in Fig.~\ref{fig:example1}(b).
In such a scenario, current approaches would select $X_2$, $A_2$ and $B_3$,
	because their QoS are optimal, resulting in a total execution time of 255 ms.
Now, if a user in France wants to execute the workflow,
	the round trip times would add over 300 ms to that time.
In comparison, executing $X_1$, $A_1$ and $B_1$ would just take 300 ms
	and only incur a minimal delay because of round trip times.
On the other hand, if providers would add the maximum delay for any user to the execution time in their respective SLAs,
	this would guarantee a certain maximum response time to all users,
		but it would also discourage users from selecting local providers
			and instead favor providers with the most homogeneous delays towards all users
				(e.g. providers in the center of Fig.~\ref{fig:example1}(b) in France).

\subsection{Service Architecture}
\begin{figure}[htb!]
	\centering
	\includegraphics[width=\twoImages]{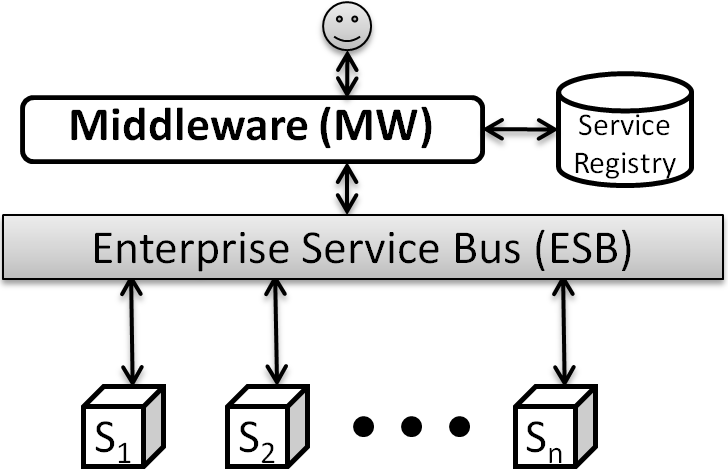}
	\caption{Standard Architecture}
	\label{fig:architecture}
\end{figure}

\noindent In fact, the standard service-oriented architecture (SOA) enforces this problem.
As shown in Fig.~\ref{fig:architecture}, 
	the actual idea of the architecture is to make the network locations transparent to the middleware,
		taking away this relevant decision information from the composition process.

\noindent In such a centralized architecture \cite{Papazoglou2003} all communication between middleware and service happens through the Enterprise Service Bus (ESB).
If we execute the workflow from Fig.~\ref{fig:example1} this means that there will be no direct communication between X and B.
Instead, the middleware will call X, wait for the result of X, and then call B, etc.,
	causing unnecessary network overhead.
While there also exist decentralized SOAs avoiding that overhead, as in the context of executing e.g. WS-CDL \cite{WS-CDL2005} service choreographies,
	the standard architectures for service orchestrations are all centralized, including the one assumed by BPEL \cite{BPEL2006}.

A simple way to solve this problem would be to design a SOA in which services can communicate their results directly to each other.
Even if we ignore the business logic of a workflow that has to be evaluated somewhere,
	there still are significant obstacles to such an architecture.
First, it would not be trivial to implement such an architecture,
	as services would have to perform several middleware functions (e.g. wait for/buffer input data, handling missing data/timeouts, etc.).
Also, there is the principle of the separation of concerns
	which tells us that a service provider might not want or should not be bothered with implementing such additional functionality.
Finally, probably the biggest obstacle is posed by the fact that current standards are already widely adopted.
Introducing new requirements for all service providers would most likely lead either to poor acceptance or to a fragmentation of the market.
Thus, we instead propose a scalable distributed service architecture that minimizes network delay and transfer times,
	while requiring no changes on the provider side, allowing for a gradual adoption.

\subsection{Network Transfer}
\begin{figure}[htb!]
	\centering
  \includegraphics[width=\oneImage]{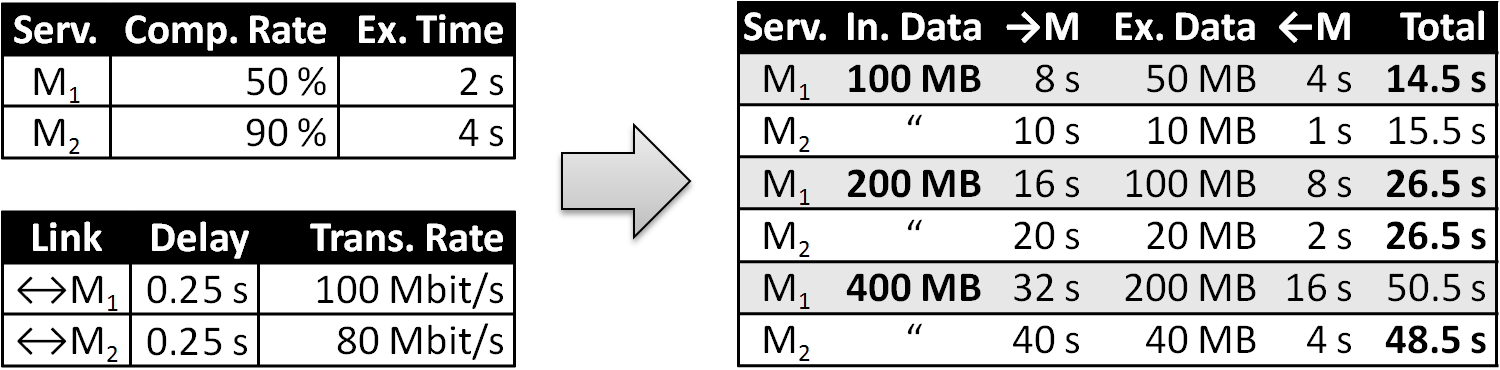}
	\caption{Audio Encoding Example}
  \label{fig:exampleTransfer}
\end{figure}

\noindent In addition to network delays,
	the transfer of data across the network can also account for a significant amount of time,
		as our example in Fig.~\ref{fig:exampleTransfer} illustrates.

\noindent Given services $M_1$ and $M_2$ that take raw audio data
	and return encoded audio data,
		their difference is the execution time (\textit{Ex. Time}) and the compression rate (\textit{Comp. Rate}) in relation to the input data.
For instance, sending 100 MB of raw audio data to $M_1$ takes 8s over a 100 MBit/s link,
	with 50 MB of resulting encoded data (\textit{Ex. Data}).
While most current approaches would probably prefer $M_1$, because of its lower execution time (and higher transfer rate),
	in practice calling $M_1$ is only faster, if we send less than 200 MB of audio data.
For more data $M_2$ is faster because of its superior compression. 
This example also shows that the QoS of a service cannot always be specified as static values in a SLA,	as it is common.
Instead a service provider might need to specify input-dependent QoS;
	especially for data-driven scenarios this can be quite significant.

\subsection{Contributions}
\noindent Thus, we present the following contributions to realize a distributed architecture for network- and QoS-aware service composition:
\begin{enumerate}
	\item A scalable \textbf{distributed service architecture} composed of a flexible number of distributed control nodes;
			it generalizes the standard architecture and 
			adapts from a completely centralized to a completely distributed realization by adding control nodes as needed.
	\item A network-aware \textbf{QoS aggregation algorithm} that allows to accurately estimate the QoS of service compositions
		executed in a distributed fashion through our architecture,	extending \cite{Klein2012}.
\end{enumerate}
Note that our architecture can be gradually adopted, as it requires no changes to existing services.
It generalizes the implicitly introduced completely distributed architecture from our previous network-aware approach \cite{Klein2012}.
The network QoS we consider are latency and transfer rates.
In our evaluation we show that our architecture is near-optimal even with a limited number of control nodes.

The structure of this paper is as follows.
Section~\ref{related} reviews related work.
Section~\ref{approach} defines our approach consisting of our architecture and QoS aggregation algorithm.
Section~\ref{evaluation} evaluates the benefits of our approach.
Finally, Section~\ref{conclusion} concludes the paper.

\section{Related Work}
\label{related}
\noindent In this section we survey related work from the following four categories.

\subsection{QoS-aware Service Composition}
\noindent The foundation for our research is given in \cite{Zeng2003} where the QoS-aware composition problem (CP) is introduced.
Common notions, which we also use, are given, and the problem is formalized and solved with (Linear) Integer Programming (IP),
	which is still a common way to obtain optimal solutions for the CP.
A genetic algorithm (GA) is used in \cite{Canfora2005,Jaeger2007}.
Besides, many efficient heuristic algorithms have been introduced in \cite{Alrifai2009,Lecue2009,Yu2007},
	and most recently in \cite{Alrifai2010,Klein2011,Rosenberg2010}.
All these approaches share the same definition of the CP which ignores the QoS of the network connecting the services.
Except for IP which requires a linear function to compute the utility of a workflow,
	most approaches can be easily augmented with our two-phased QoS algorithm.

\subsection{Advanced QoS}
\noindent The previously mentioned approaches all simply aggregate static QoS values defined in SLAs.
Time-dependent QoS evaluated depending on the execution time are given in \cite{Kloepper2010}.
As we will see, our algorithm computes when the execution of each service starts,
	so we can also compute time-dependent QoS.
SLAs with conditionally defined QoS are given in \cite{Klein2009},
	which can be considered a special case of input-dependent QoS,
		and, thus, can be handled by our approach, as well. 

In \cite{Menasce2009} constraints on the choice of providers are given,
	requiring certain services to be executed on the same provider.
Introducing such constraints for critical services could also reduce network delay and transfer times to some extent.
This would require a significant effort to introduce such heuristic constraints though,
	while still not necessarily leading to a (near-)optimal solution.	

\subsection{Network QoS}
\noindent Many approaches, such as \cite{Boutaba2005,Jin2007}, deal with point-to-point network QoS,
	but they do not consider services and compositions from SOC.
One of the few examples that combines this with SOC is \cite{Ye2011}
	which looks at service compositions in cloud computing.
The difference is that instead of the normal composition problem a scheduling problem is solved where services can be deployed on virtual machines at will.
Also no QoS algorithm is given, so it is unclear, if that approach can compute input-dependent QoS and network transfer times.

\subsection{Workflow Scheduling}

\noindent In the related field of workflow scheduling,
	a workflow is mapped to heterogeneous resources (CPUs, virtual machines, etc.),
		and information about the network is sometimes considered, as well.
The goal is to achieve a \mbox{(near-)}\\optimal scheduling minimizing the execution time,
	which is often achieved by greedy heuristic approaches, like HEFT \cite{To2002}.
The reason such greedy algorithms seem to suffice is that only one QoS property (response time) is optimized,
	and that no QoS constraints have to be adhered to, greatly simplifying the problem.
Thus, while the setting is similar to ours,
	the complexity of the problem is quite different,
		as we optimize multiple QoS properties under given QoS constraints.

\section{Approach}
\label{approach}
\noindent In this section we define our approach.
First, we present our proposed architecture.
Based on that, we explain and motivate our workflow model.
Then, we define our execution policy.
Finally, we describe our algorithmic framework to compute the QoS of a workflow.

\subsection{Distributed Service Architecture}
\begin{figure}[htb!]
	\centering
	\includegraphics[width=\twoImages]{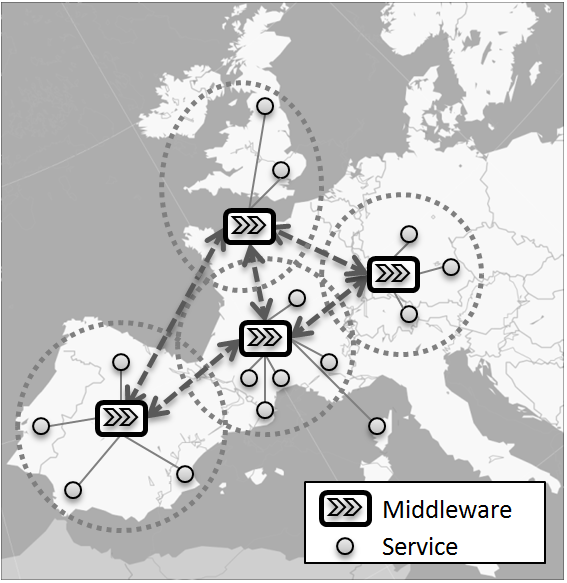}
	\caption{Distributed Middleware}
	\label{fig:distributed_middleware}
\end{figure}

\noindent We want to minimize communication costs with a distributed middleware	that could be deployed as in Fig.~\ref{fig:distributed_middleware}.

\noindent While not requiring services to communicate directly with each other,
	we can still save network costs by delegating to call services to a part of our middleware $M_{local}$ which is closer to them.
The more places exist where we can deploy such a part of our middleware the better,
	but even just a few places would allow us to reduce the communication cost significantly.

\begin{figure}[htb!]
  \centering
	\includegraphics[width=\oneImage]{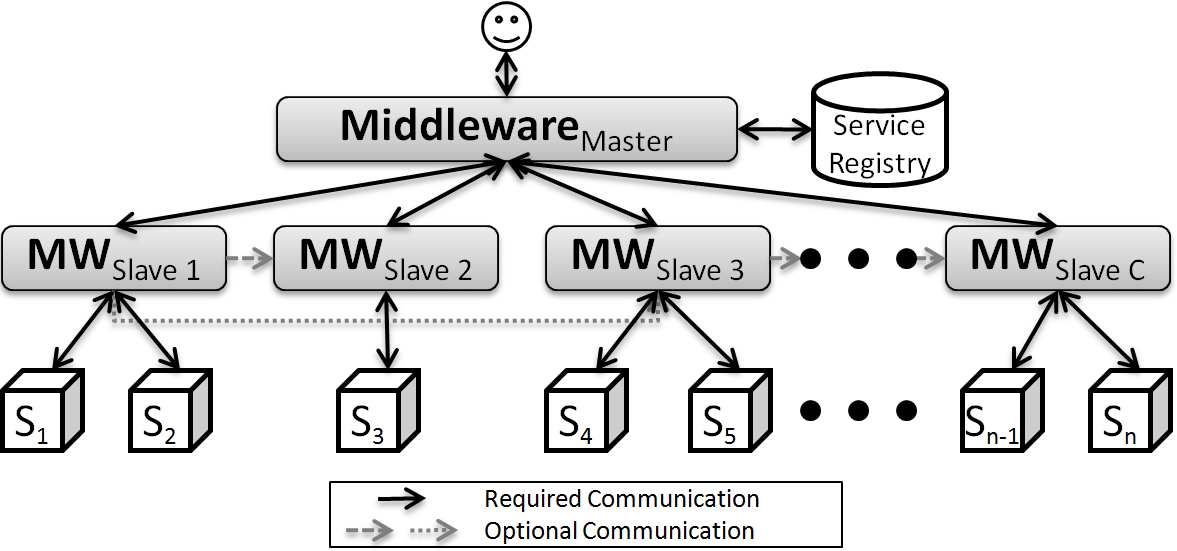}
	\caption{Distributed Architecture}
	\label{fig:distributed_architecture}
\end{figure}

While there are many ways to distribute the middleware,
	we propose the customized master-slave pattern depicted in Fig.~\ref{fig:distributed_architecture}
		because of its simplicity and robustness.
In this architecture, the master control node performs the bulk of the middleware tasks
	such as discovery, selection, fault-handling, etc.
The only thing that is delegated to the slave control nodes is executing nearby services,
	and the collection and transmission of their results.
Information necessary for fault-handling, monitoring and other tasks is forwarded to the master control node
	which makes any necessary decisions.

A slave control node basically only has to know which services it needs to execute.
Additionally, it waits for necessary data to arrive before the execution,
	and afterwards sends the obtained results to all the slave and/or the master control nodes as needed.
We will give an execution policy which achieves this in Sec.~\ref{exPolicy}.
Of course, the master control node itself can also execute services, e.g. if they are close,
	or if no closer slave control node can be deployed.
Thus, in the case of no slave control nodes,	this architecture is equivalent to the standard service architecture.
If a slave control node can be deployed at any network location,
	our architecture is equivalent to a maximally distributed architecture.
Note that we would assume the number of these network locations to be limited at least in the near future.

\subsection{Workflow Model}
\noindent We now give the workflow model that provides the basis of our approach.
We first introduce our common node concept, before describing our model.

\subsubsection{Node Concept}
\noindent While we mostly adhere to common notations, there is one main difference:
We consider an executable workflow to consist of nodes,
	whether be they logical nodes or service nodes.
Service nodes represent traditional tasks, and logical nodes represent business logic,
	such as loops, conditions, etc.
As we distribute our middleware,
	we can choose the network locations of both node types according to available options.
Both node types have incoming and outgoing links which inhibit a certain QoS,
	and, thus, we consider them both equally in our model and in our computations.

\subsubsection{Model Representation}
\noindent We support the common workflow patterns of sequences, parallel invocations (alternative, parallel, etc.) and loops,
	as in \cite{Jaeger2004}.
Logical nodes must be explicit in the model, as they have to be distributed, as well.
Model-wise the corresponding structure of our workflows is defined as:

\begin{equation*}
	\begin{array}{lcl}
	  \displaystyle
	  	W = \left \{
				\begin{array}{clr}
					S & \mbox{Service Node}\\
					L & \mbox{Logical Node}\\
					P(w_1, ..., w_n) & \mbox{Workflow. Pattern} \\
				\end{array}
			\right .
			\\
			\mbox{with } P \in \left\{\mbox{Seq, AND, XOR, OR, Loop} \right\}
	\end{array}
	\label{eq:doesExceed}
\end{equation*}

\begin{figure}[htb!]
	\centering
	\includegraphics[width=\oneImage]{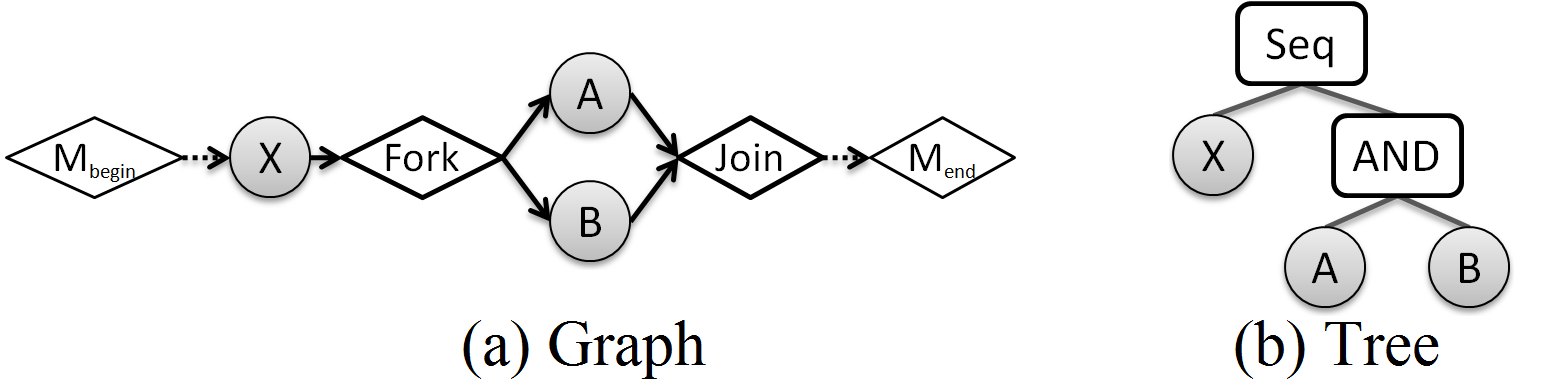}
	\caption{Model Representations of an Example Workflow}
	\label{fig:representations}
\end{figure}

\noindent For instance, the model of our previous example workflow is depicted in Fig.~\ref{fig:representations}(a).
We added explicit fork and join nodes that perform the necessary processing to call A and B in parallel and to join the results afterwards.
Also we introduced explicit start and end nodes which both correspond to the master of the middleware
	in order to compute the correct total network QoS.
	
\noindent Such workflows are commonly given in a hierarchical manner
  as a tree representation like in Fig.~\ref{fig:representations}(b), e.g. if specified in BPEL~\cite{BPEL2006}.
We first have to convert this tree representation to a graph, before we can compute the QoS of the workflow,
	and add explicit logical nodes.
For that purpose we need two helper functions
	to compute the predecessors and successors of a node in a workflow.
The function $first(wf)$, given in the following, computes all atomic nodes of a (sub) workflow $wf$ that are executed first (within that workflow).
In an analog way, $last(wf)$ computes the nodes that will be executed last.

\begin{algorithm}
	\label{algo:first}
	\begin{algorithmic}[1]
		\Procedure{first}{$wf$}
			\If{$wf = S \| L$}
				\State \textbf{return} \{$wf$\}
			\ElsIf{$wf = Seq \| Loop (w_1,...,w_n)$}
				\State \textbf{return} first($w_1$)
			\ElsIf{$wf = AND \| XOR \| OR (w_1,...,w_n)$}
				\State \textbf{return} first($w_1$)
			\EndIf
		\EndProcedure
	\end{algorithmic}
\end{algorithm}

\noindent Using these functions, we can convert a hierarchical workflow into a directed graph with the following $mapToGraph$ algorithm.
The algorithm structurally traverses the hierarchical structure of the workflow 
	in a depth-first manner until it finds an atomic service node which can be connected with its preceding and succeeding service nodes.
Once we have converted a workflow into a directed graph we can compute the QoS of the workflow.

\begin{algorithm}
	\label{algo:mapToGraph}
	\begin{algorithmic}[1]
		\Procedure{mapToGraph}{$fs, wf, ls, g$}
			\If{$wf = S \| L$}
				\State $\{\forall f \in fs \; . \; \mbox{\textbf{add} edge } (f \rightarrow wf) \mbox{ to } g \}$
	    	\State $\{\forall l \in ls \; . \; \mbox{\textbf{add} edge } (wf \rightarrow l) \mbox{ to } g \}$
			\ElsIf{$wf = Seq \| Loop (w)$}
				\State \textbf{mapToGraph}($fs, w, ls, g$)
			\ElsIf{$wf = Seq \| Loop (w_1,...,w_n)$}
				\State $h = w_1$, $t = w_2,...,w_n$
	      \State \textbf{mapToGraph}($fs$, $h$, first(Seq/Loop($t$)), $g$)
	      \State \textbf{mapToGraph}(last($h$), Seq/Loop($t$), $ls$, $g$)
			\ElsIf{$wf = AND \| XOR \| OR (w_1,...,w_n)$}
				\State $\{\forall i \in \{1..n\} \; . \; $\textbf{mapToGraph}($fs, w_i, ls, g$)$ \}$
			\EndIf
		\EndProcedure
	\end{algorithmic}
\end{algorithm}

\subsection{Execution Policy}
\label{exPolicy}
\begin{figure}[hb!]
	\centering
	\includegraphics[width=\oneImage]{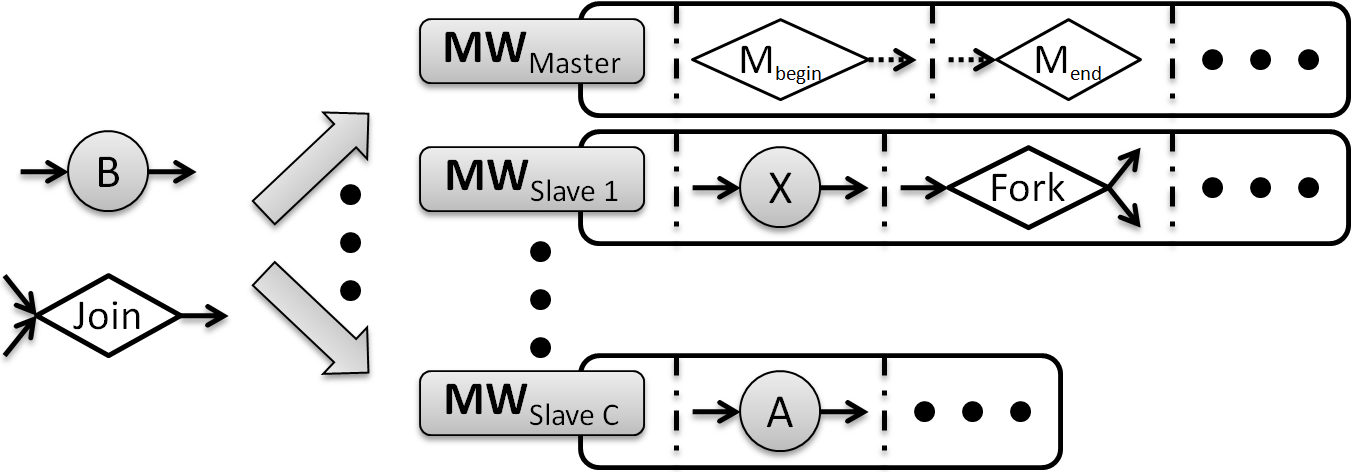}
	\caption{Distributed Workflow Execution}
	\label{fig:workflow_execution}
\end{figure}

\noindent Before computing the QoS of a workflow,
	we have to define the execution policy.
As mentioned before, we assume that our middleware is distributed.
Our main goal is to minimize the amount of knowledge and processing required of our slaves.
Thus, we propose the execution policy shown in Fig.~\ref{fig:workflow_execution}.
The main work is done by the middleware master.
After the master has determined the optimal services and slaves,
	each slave gets deployed and work packages are distributed to the slaves as in Fig.~\ref{fig:workflow_execution}.
Each work package contains exactly one service node, plus information about preceding and succeeding service nodes.

Once a slave has received all results from the preceding service nodes,
	it can execute its service node(s), e.g. call the service or evaluate the business logic,
		and send the result to all succeeding service nodes.
The sending and receiving is handled by the corresponding slave or master.
The final result is returned to the master,
	but intermediate results are just passed on as needed (e.g. they might never pass through the master).
Thus, a slave does not know about the structure of the workflow or a part of it, it just executes work packages.
If more complex decisions have to be made,
	for example on a service failure,
		then the slave reports all available information to the master
			which performs the necessary rescheduling.
      
\subsection{QoS Computation}
\label{qos:formal}

\noindent In order to compute the execution duration of a workflow,
	we simulate its execution with the $simulateExecution$ algorithm
		introduced in the following.
Note that the commonly used aggregation is not sufficient to compute this,	as we will illustrate later.
For each service node of the workflow, we keep track of how many preceding service nodes still need to be executed (line 4).
Then, we execute ready nodes (line 6) until no nodes are left.
We also keep track of the time when the execution of a node has started and finished.
After a service node is executed, we evaluate its QoS (line 8), e.g. according to its SLA,
	and then virtually pass its result to all succeeding nodes (line 10).

\begin{algorithm}
	\label{algo:simulateExecution}
	\begin{algorithmic}[1]
		\Procedure{simulateExecution}{$g$}
			\For{each vertex $v \in g$}
				\State v.execStart = 0
				\State v.reqIn = $\left|v.incoming\right|$
			\EndFor
			\While{$\exists \mbox{ unvisited } v \in g \; . \; v.reqIn = 0$}
				\State \textbf{visit} any unvisited $v \in g \; . \; v.reqIn = 0$
				\State evaluateQoS($v$)
				\State v.execEnd = $v.execStart + v.qos.runtime$
				\For{each $w \in v.outgoing$}
					\State cv = $v.controlNode$
					\State cw = $w.controlNode$
					\State cNet = getNetworkQoS($cv, cw$)
					\State vNet = getNetworkQoS($v, cv$)
					\State wNet = getNetworkQoS($w, cw$)
					\State trans = $\displaystyle\sum_{\substack{net \; \in \\\{cNet, vNet, wNet\}}}{\frac{v.resultSize}{net.transRate}}$
					\State delay = $\displaystyle\sum_{\substack{net \; \in \\\{cNet, vNet, wNet\}}}{net.delay}$
					\State end = $v.execEnd + trans + delay$
					\State w.execStart = $\textbf{max}\{w.execStart, end\}$
					\State w.reqIn -= 1
				\EndFor
			\EndWhile
		\EndProcedure
	\end{algorithmic}
\end{algorithm}

\noindent When passing the result we consider the delay and the duration of the data transfer between the two service nodes (line 18).
For instance if a service node $v$ needs to communicate with a service node $w$,
	then first $v$ communicates with its control node $cv$,
		$cv$ communicates with $w$'s control node $cw$,
			and finally $cw$ communicates with $w$.
Accordingly, we compute both the network transfer time (line 16) and the network delay (line 17).
Note that in order to estimate the data transfer we either need some SLA specifying that, or some historical data.

\begin{figure}[htb!]
	\centering
	\includegraphics[width=\oneHalfImages]{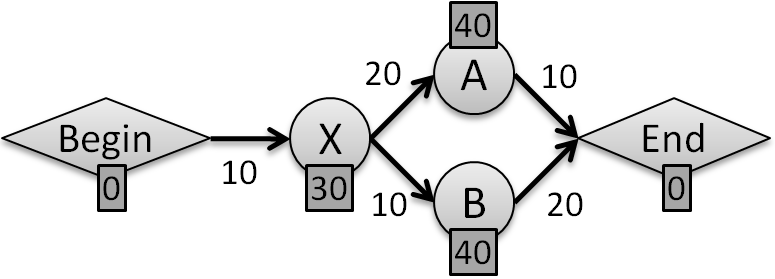}
	\caption{QoS of a Workflow}
	\label{fig:workflow_qos}
\end{figure}

\begin{figure}[htb!]
	\centering
  \includegraphics[width=\oneHalfImages]{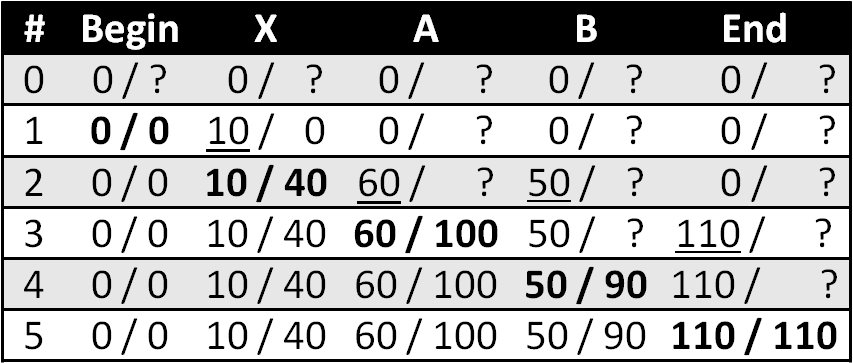}
	\caption{Simulated QoS}
	\label{fig:workflow_qos_times}
\end{figure}

If we annotate our previous workflow example with execution durations of the services,
	and network delays, as in Fig.~\ref{fig:workflow_qos},
		our algorithm will produce the values of Fig.~\ref{fig:workflow_qos_times} for the execution times of the nodes ($start/end$).
This simple example shows that hierarchical QoS aggregation would not work,
	because first A and B would be aggregated together.
This would make it impossible to compute the correct network QoS,
	as the maximum of the delay of incoming and outgoing nodes of (A,B) each would be aggregated as 20.
But actually there exists no path that can go through both the incoming and the outgoing nodes of (A,B) with delay 20,
	so that the aggregated value would be too high.
	
\noindent Thus, we compute the QoS with a two phased algorithm as in \cite{Klein2012}.
In the first phase, we simulate the execution of the workflow with $simulateExecution$
	based on the graph we obtain by applying $mapToGraph$.
In the second phase, we take the obtained QoS for each node and aggregate it in a hierarchical manner
	according to the commonly used aggregation rules that take our workflow patterns into account, as in \cite{Jaeger2004,Yu2007}.
Just for the runtime of the workflow, we keep the computation from the first phase,
	as we cannot compute it with a hierarchical aggregation, like argued previously.

\section{Evaluation}
\label{evaluation}
\noindent In this section we evaluate our approach.
First, we describe the setup of our evaluation.
Then,	we evaluate the benefits of our architecture.
Finally, we show that our architecture also scales in regard to the problem size.

\begin{figure}[b]
	\centering
	\includegraphics[width=\oneImage]{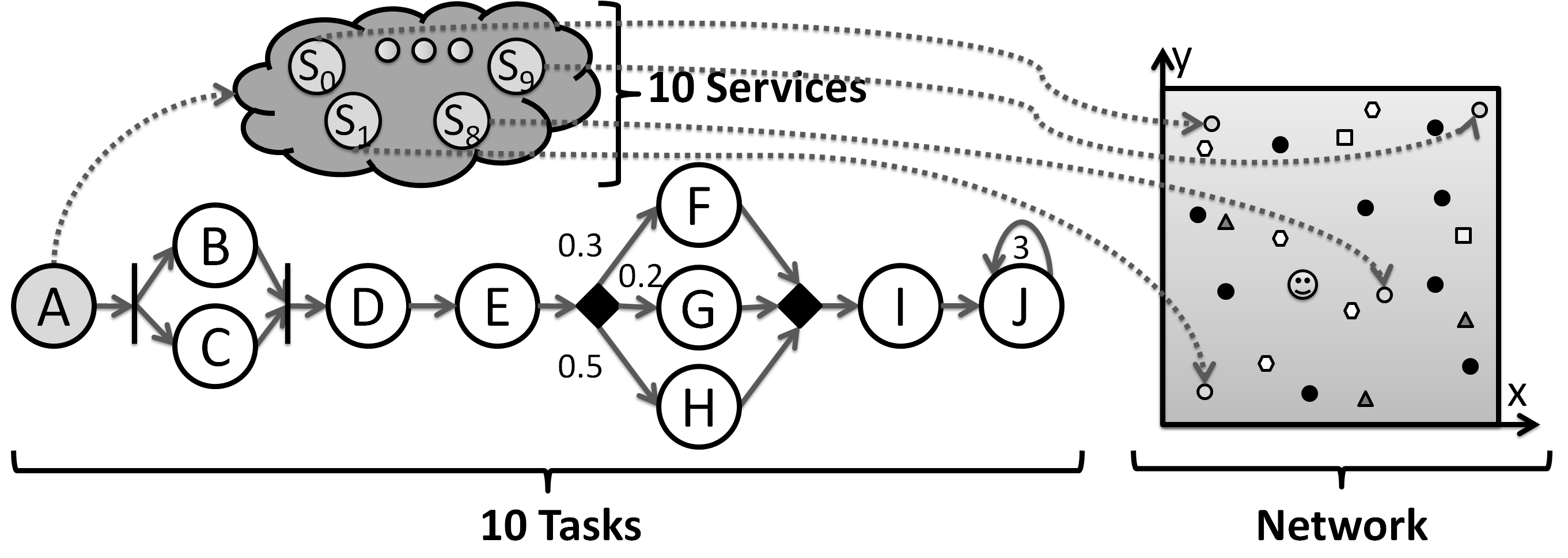}
	\caption{Example Workflow of Size 10}
	\label{fig:generated_workflows}
\end{figure}

\subsection{Setup}
\noindent The evaluation was run on a machine with 32 AMD Opteron cores with 2.4 GHz.
All algorithm instances were evaluated in separated threads and granted a single exclusive core,
	while memory was shared and less than 1 GB per instance was needed.
Note that the following evaluations settings are based on our previous evaluations in \cite{Klein2012}.
	
We generated 100,000 unique network locations.
The workflows were generated with random tasks and control structures.
For each task, we randomly chose a number of network locations and created services there.
Figure~\ref{fig:generated_workflows} depicts an example of a generated workflow of length 10.
We chose the following algorithms which all use our new QoS aggregation algorithm.
\begin{itemize}
	\item \textbf{Dijkstra}, an optimal algorithm for the shortest-path problem.
	\item \textbf{GA*}, a standard genetic algorithm with population of size 100.
	\item \textbf{NetGA}, our previous network-aware approach introduced in \cite{Klein2012}. The size of the population is 100, as well. 
\end{itemize}
In addition, we evaluated optimal variations of those algorithms, marked with "[o]",
	which could deploy an unlimited number of control nodes.

\subsection{Latency}
\begin{figure}[t]
	\centering
	\includegraphics[width=\oneImage]{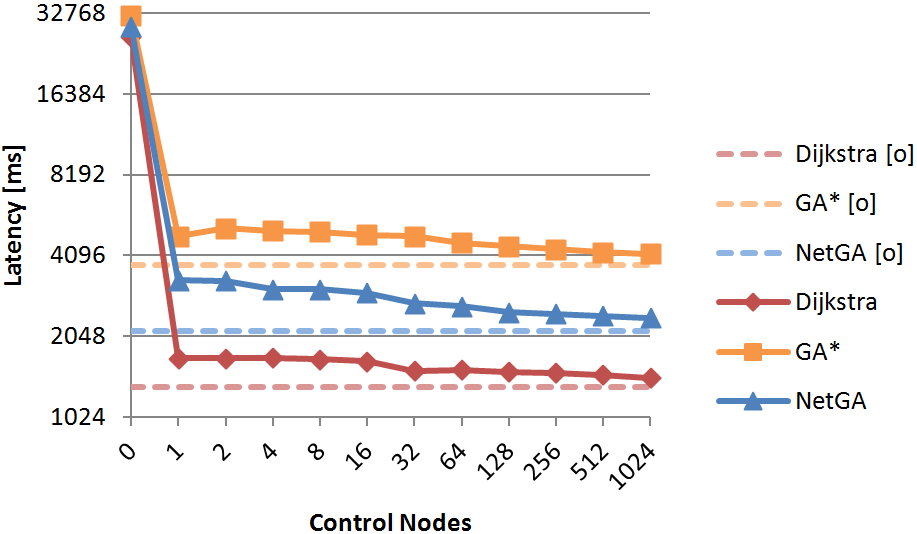}
	\caption{Latency vs. Number of Control Nodes\\(Workflow Size 40)}
	\label{fig:eval_lat}
\end{figure}

\noindent By adjusting the number of control nodes,
	our architecture can adapt from being completely centralized (no control nodes)
		to being completely distributed (unlimited number of control nodes).
Note that, strictly speaking, by control nodes we mean additional (slave) control nodes,
	as we always have one (master) control node run by the user requesting the workflow's execution.
Figure~\ref{fig:eval_lat} plots the latency of found service compositions against the number of control nodes,
	with a fixed workflow size of 40.
The control nodes were chosen randomly.
We can observe the following two things.
First, a completely centralized architecture results in a quite bad latency for the service compositions evaluated.
Second, once a sufficient number of control nodes ($\geq 32$) is deployed,
	the latencies of the algorithms come reasonably close to their optimal variations.
Furthermore,
	using 1024 control nodes, already results in near optimal results for all algorithms in our experiments.
Note that, as shown in \cite{Klein2012}, our \textbf{NetGA} algorithm is better in approximating \textbf{Dijkstra} than the standard genetic algorithm, \textbf{GA*};
	also \textbf{Dijkstra} is used only for comparison purposes as it cannot be applied to many common service scenarios
		and as its performance does not scale well in realistic settings.
    
\subsection{Scalability}
\begin{figure}[t]
	\centering
	\includegraphics[width=\oneImage]{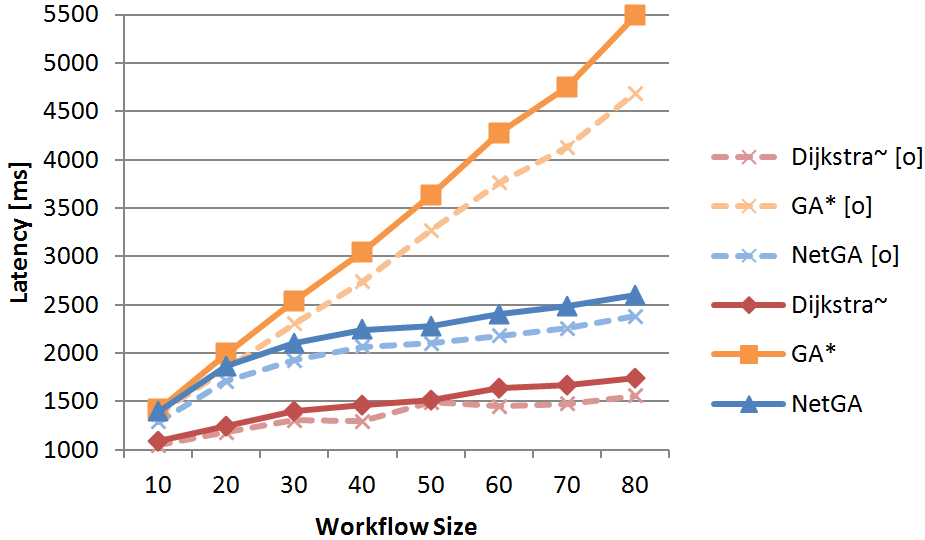}
	\caption{Latency vs. Workflow Size\\(1024 Control Nodes)}
	\label{fig:eval_wf}
\end{figure}

\noindent Our final evaluation, plotted in Fig.~\ref{fig:eval_wf}, shows how well our architecture scales in term of the problem size.
For a fixed number of 1024 control nodes,
	we can see that the optimality of our \textbf{NetGA} algorithm does not decrease significantly for the workflow sizes ($\leq 80$) we evaluated.
Thus, even for more complex settings we do not have to increase the number of control nodes.
Note that the optimality of \textbf{GA*} seems to decrease slightly,
	but, as mentioned in \cite{Klein2012}, \textbf{GA*} is not very efficient at optimizing the latency in the first place.

\section{Conclusion}
\label{conclusion}
\noindent In this paper we motivated the network- and QoS-aware service composition problem
	which is highly relevant in today's distributed environments.
Then, we introduced a scalable distributed service architecture which significantly reduces network delay and transfer times
	by eliminating unnecessary communication required in case of a central middleware, as it is common today.
We showed that our architecture is a generalization of the standard architecture,
	being able to adopt from a completely centralized to a completely distributed realization.
As it requires no changes on the side of service providers,
	it guarantees compatibility to existing services and allows for gradual adoption.
On top of that, we introduced an extended QoS aggregation algorithm
	that estimates real-world QoS performance by computing the network QoS for any realization of our architecture.
Our algorithm can easily be used to augment current approaches.
Finally, we evaluated the benefits of our architecture,
	showing that it is near-optimal even with a limited number of control nodes.

As future work we want to explore algorithms for choosing good control nodes.
In this work we just selected the control nodes randomly, as it did not make a significant difference how we picked those nodes
	except when just choosing one or two control nodes.
We think that in realistic settings there are many factors,
	such as availability or queuing times, that could effect how useful certain control nodes are.
Also minimizing the number of control nodes even further while still obtaining near-optimal results could be critical in certain settings.

\section{Acknowledgments}
\noindent We would like to thank Florian Wagner for the detailed feedback that helped us to improve our approach.
Also we would like to thank Michael Nett and Michael E. Houle at the National Institute of Informatics (Japan) for kindly offering us one of their machines for running our extensive evaluations.
Adrian Klein is supported by a Research Fellowship for Young Scientists from the Japan Society for the Promotion of Science.

\bibliographystyle{IEEEtranS}
\bibliography{net_arch_ext}

\end{document}